# On Big Data Benchmarking

[1]Rui Han and [2]Xiaoyi Lu

[1]Department of Computing, Imperial College London
[2]Ohio State University

r.han10@imperial.ac.uk, luxi@cse.ohio-state.edu

**Abstract**

Big data systems address the challenges of capturing, storing, managing, analyzing, and visualizing big data. Within this context, developing benchmarks to evaluate and compare big data systems has become an active topic for both research and industry communities. To date, most of the state-of-the-art big data benchmarks are designed for specific types of systems. Based on our experience, however, we argue that considering the complexity, diversity, and rapid evolution of big data systems, for the sake of fairness, big data benchmarks must include diversity of data and workloads. Given this motivation, in this paper, we first propose the key requirements and challenges in developing big data benchmarks from the perspectives of generating data with 4V properties (i.e. volume, velocity, variety and veracity) of big data, as well as generating tests with comprehensive workloads for big data systems. We then present the methodology on big data benchmarking designed to address these challenges. Next, the state-of-the-art are summarized and compared, following by our vision for future research directions.

*Categories and Subject Descriptors* D.2.8 [**Software Engineering**]: Metrics – performance measures

*General Terms* Measurement, Performance.

*Keywords* big data systems; benchmark; data; tests

## 1. Introduction

Big data systems have gained unquestionable success in recent years and will continue its rapid development over the next decade. These systems cover many industrial and public service areas such as search engine, social network and e-commerce, as well as a variety of scientific research areas such as bioinformatics, environment, meteorology, and complex simulations of physics. Conceptually, big data are characterized by very large data volumes and velocities, diversity and variety (various data types, and complex data processing requirements. In the era of big data, these data require a new generation of big systems to capture, store, search, and analyze them within an acceptable elapsed time. The complexity, diversity, and rapid evolution of big data systems give rise to new challenges in how to compare their performance, energy efficiency, and cost effectiveness.

Conceptually, a big data benchmark aims to generate *application-specific* workloads and tests capable of processing big data sets to produce meaningful evaluation results [18]. Considering the diversity of big data systems (e.g. there are three mainstream application domains, namely search engine, social network, and e-commerce, of internet service workloads), and the emergence of new systems driven by the exploration of big data value, covering diversity of workloads is the prerequisite to perform successful and efficient benchmarking tests. Within this context, we propose our insights into the requirements and challenges in developing big data benchmarks. We also present the methodology on big data benchmarking, which represents our thinking about how to address these challenges. Finally, we discuss state-of-the-art benchmarking techniques for big data systems and propose some future research directions. The aim of this paper, therefore, is to provide a foundation towards building a successful big data benchmark, and to stimulate productive thinking, investigation, and development in this research area.

## 2. Requirements and Challenges

Big data benchmarks are developed to evaluate and compare the performance of big data systems and architectures. Figure 1 shows the benchmarking process for big data systems that consists of five steps. At the *Planning* step, the benchmarking object, application domain, and evaluation metrics are determined. In the following two steps, the data and test used in evaluation are generated. Next, the benchmark test is conducted at the *Execution* step and the evaluation result is reported. At the last step, the benchmarking result is analysed and evaluated.

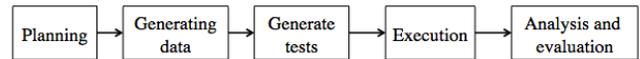

**Figure 1. Benchmarking process for big data systems**

Successful and efficient benchmarking can provide realistic and accurate measuring of big data systems and thereby addressing two objectives. (1) Promoting the development of big data technology, i.e. developing new architectures (processors, memory systems, and network systems), innovative theories, algorithms, techniques, and software stacks to manage big data and extract their value and hidden knowledge. (2) Assisting system owners to make decisions for planning system features, tuning system configurations, validating deployment strategies, and conducting other efforts to improve systems. For example, benchmarking results can identify the performance bottlenecks in big data systems, thus optimizing system configuration and resource allocation. In this section, we present the requirements and challenges in building big data benchmarks.

### 2.1 Generating Data with the 4V Properties of Big Data

Technically, a big data system can be attributed four dimensions: volume, velocity, variety, and veracity, which form the "4V" properties of big data [5]. (1) Volume represents the amount/size of data such as Terabyte (TB) or Petabyte (PB). (2) Velocity reflects the speed of generating, updating, or processing data. (3) Variety denotes the range of data types and sources. Finally, (4) veracity reflects whether the data used in benchmarking conform to the inherent and important characteristics of raw data.

In a big data benchmark, applying real-world data or generating synthetic data for application-specific workloads is a central problem. Traditionally, although some benchmarks use real data sets as inputs of their workloads and thereby guaranteeing the data veracity, the volume and velocity of real data sets cannot be flexibly adapted to different benchmarking requirements. Based on our experience, we also noticed that in many practical scenarios, obtaining a variety of real data is not trivial because many data owners are not willing to share their data due to confidential issues. In big data benchmarks, therefore, the consensus is to generate synthetic data as inputs of workloads on the basis of real data sets. Hence in synthetic data generation, preserving the 4V properties of big data is the foundation of producing meaningful and credible evaluation results.

**Volume**. Today, data are generated faster than ever. For example, about 2.5 quintillion bytes of data are created every day [5] and this speed is expected to increase exponentially over the next decade according to International Data Corporation (IDC). In Facebook, there are 350 million photos updated and more than 500 TB data generated per day. The above facts indicate the big data generators must be able to generate different volumes of data as inputs of typical workloads. The data volume has different meanings in different workloads. For example, in workloads for processing text data (e.g. *sort* or *WordCount*), the volume is represented by the amount of data (e.g. 1 TB or 1 PB text data). In social network graph workloads, the volume is represented by the number of vertices in social graph data (e.g. $2^{20}$ vertices).

**Velocity.** In the context of big data generation, data velocity has threefold meanings. First of all, it represents the *data generation rate*. For example, generating 100 TB text data in 10 hours means the generation rate is 10 TB/hour. Secondly, many big data applications have real-time data updating; that is, data velocity represents the *data updating frequency* in this case. For example, in a social network site, the social graph data are continuously updating. Finally, in streaming processing systems, data streams continuously arrives and these streams must be processed in real-time to keep up with their arriving speed. Hence data velocity represents the processing speed. Given the above facts, it is challenging to reflect data generation rates, updating frequencies, and processing speeds in data generation.

**Variety**. The fast development of big data systems gives birth to a diversity of data types, which cover structured data (e.g. tables), unstructured data (e.g. text, graph, images, audios, and videos), and semi-structured data (e.g. web logs, reviews, and resumes, where reviews and resumes contains both text and graph data). Hence in big data benchmarking, it is required that the data generators can support the whole spectrum of data types including structured, semi-structured, and unstructured data, as well as representative data sources such as table, text, stream, and graph. It is also required that these data generators can support the complexity and diversity of workloads and keep in pace with their frequent changes.

**Veracity**. Preserving data veracity is probably the most difficult job in synthetic data generation. In big data generators, it is challenging to scale up or down a synthetic data set while keeping this data similar to the real data [18]; that is, the important characteristics of raw data must be preserved in synthetic data generation. Data veracity is important to guarantee the reality and credibility of benchmarking results.

## 2.2 Generating Benchmarking Tests with Comprehensive Workloads

Margo Seltzer et al. pointed out that a testing result is meaningful only when applying an application-specific benchmarking test [18]. Also, the diversity and rapid evaluation of big data systems means it is challenging to develop big data benchmarks to reflect various workload cases. Hence in big data benchmarks, identifying the typical workload behaviours for an application domain is the prerequisite of evaluating big data systems. Furthermore, big data benchmarks must consider the diversity of workloads to cover different types of application domains, as well as automatically generate tests based on these workloads. We now discuss the key challenges in generating workloads and tests to evaluate big data systems from the *functional view* and the *system view*.

**Functional view**.

Given the complexity and diversity of workload behaviours in current big data systems, it is reasonable to say that no single set of behaviors is representative for all applications. Hence, it is necessary to abstract from the behaviors of different workloads to a general approach. This approach should identify typical workload behaviours in representative application domains. From the functional view, these behaviors represent the system-independent outcome of processing data, thus allowing the comparison of systems of different types, e.g. a DBMS and a MapReduce system.

There are two challenges in developing this abstraction approach. First, the *operations* to process big data in a specific application domain need to be abstracted and their functions need to be identified. For example, *select*, *put*, *get*, and *delete* are abstracted operations in database systems to operate table data. Secondly, given a set of abstracted operations, *workload patterns* need to be abstracted to describe complex processing tasks by combining abstracted operations. One abstracted workload pattern can contain one or multiple abstract operations as well as their workflow. For example, an abstract pattern of *a SQL query* can contain *select* and *put* operations, in which the *select* operation executes first.

**System view**.

The abstracted operations and patterns are designed to capture the system-independent user behaviours of workloads, i.e. the data processing operations and their sequences. Thus, an abstracted benchmark test can be constructed based on abstracted operations and patterns, and this test is independent of underlying systems and software stacks. From the system view, this abstract test can be implemented over different systems and thereby allows the comparison of systems of the same type. For example, an abstract test consisting of a sequence of *read*, *write*, and *update* operations can be used to compare different DBMSs.

### 2.3 Execution

To perform a fair, efficient, and successful benchmarking test, there are several requirements and challenges to be addressed at the *Execution* step.

**Adapting to different data formats**. Since the same type of data can be stored in multiple formats, e.g. texts can be stored in a simple text file or more complex formats as web pages and pdf, Big data benchmarks need to provide format conversion, which can transfer a data set into an appropriate format capable of being used as the input of a test running on a specific system.

**Portability to representative software stacks**. A software stack consists of a set of programs working together to provide a fully functional solution. Big applications and systems belonging to one application domain are built on the basis of one or multiple software stacks. Hence covering a broad spectrum of representative software stacks in benchmark tests, as well as avoiding being too costly or difficult to port a test to a specific software stack is another important issue we need to consider.

**Extensibility**. The fast evolution of big data systems requires big data benchmarks not only keeping in pace with state-of-the-art techniques and underlying systems, but also taking their future changes into consideration. That is, big data benchmarks should be able to add new workloads or data sets with little or no change to the underlying algorithms and functions [19].

**Usability**. Usability reflects users' experiences in using benchmarks and it is a combination of factors. In big data benchmarks, these factors include ease of deploying, configuring, and use; high benchmarking efficiencies; simple and understandable performance metrics; convenient user interfaces; and so on.

## 3. On Benchmarking Methodology

### 3.1 Layer Design of Big Data Benchmarks

Figure 2 shows the layered design of a big data benchmark with three layers:

The *User Interface Layer* provides interfaces to assist system owners to specify their benchmarking requirements, such as the selected data, workloads, metrics and the preferred data volume and velocity.

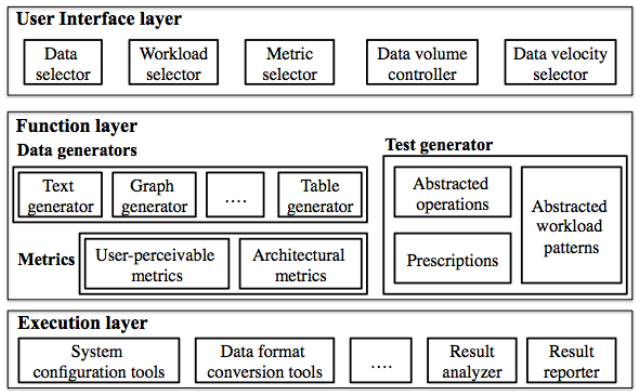

**Figure 2. Layered architecture of big data benchmarks**

The *Function Layer* has three components: data generators, test generators and metrics. Briefly, data generators are designed to produce data sets covering different data types and application domains while keeping the 4V properties of big data in these data sets. The test generator enables the automatic generation of tests with comprehensive workloads for big data systems. Metrics (either single or multiple metrics) can be divided into two types: user-perceivable metrics and architecture metrics [19]. User-perceivable metrics represent the metrics that matter for users; these metrics are usually observable and easy to be understood by users. Examples of user-perceivable metrics are the duration of a test, request latency, and throughput. While user-perceivable metrics are used to compare performances of workloads of the same category, architecture metrics are designed to compare workloads from different categories. Examples of architecture metrics are million instructions per second (MIPS) and million floating-point operations per second (MFLOPS). In addition, these metrics should not only measure system performance, but also take energy consumption, cost efficiency into consideration.

The *Execution Layer* offers several functions to support the execution of benchmark tests over different software stacks. Specifically, the system configuration tools enable a generated test running in a specify software stack. The data format conversion tools transform a generated data set into a format capable of being used by this test. The result analyzer and reporter display evaluation results.

### 3.2 Data Generators in Big Data Benchmarks

Data generators in big data benchmarks aim to efficiently generate data sets while preserving the 4V properties of big data [14]. Figure 3 shows the process of generating data sets.

At the first step, data generators support the *variety* of big data by selecting real data sets to cover representative application domains as well as different data sources and types. The generators can also apply tools to directly generate synthetic data sets; that is, these synthetic data sets are independent of real data. This is because it is accepted that such purely synthetic data can be used as inputs of some workloads such as the *Sort* and *WordCount* workloads in Micro benchmarks; and the *Read*, *Write*, and *Scan* workloads belonging to basic database operations.

At the second step, each data generator employs a data model to capture and preserve the important characteristics in one or multiple real data sets of a specific date type. For example, a text generator can apply Latent dirichlet allocation (LDA) [8] to describe the topic and word distributions in text data. This generator first learns from a real text data set to obtain a word dictionary. It then trains the parameters $\alpha$ and $\beta$ of a LDA model using this data set. Finally, it generates synthetic text data using the trained LDA model. To preserve data veracity, it is required that different models should be developed to capture the characteristic of real data of different types such as table, text, stream, and graph data. In addition, the sampling tools enable the scaling down of data set sizes.

At the third step, the volume and velocity can be controlled according to user requirements. For example, the data generation can be paralleled and distributed to multiple machines, thus supporting different data generation rates.

At the fourth step, after a data set is generated, the format conversion tools transform this data set into a format capable of being used as the input data of a specific workload.

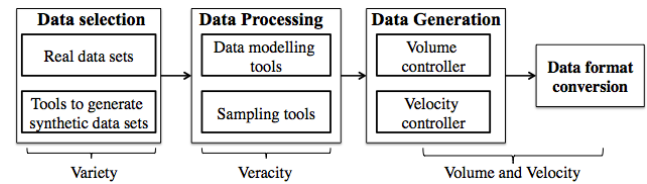

**Figure 3. The big data generation process**

### 3.3 Test Generator in Big Data Benchmarks

In big data benchmarks, the test generator is developed to automatically generate benchmarking tests for big data systems. The basic idea of this generator is to abstract from the workload behaviours of current big data systems to a set of *operations* and *workload patterns* used in big data processing [20]. As shown in Figure 2, the test generator consists of three components.

**Operations** represent the abstracted processing actions (operators) on data sets. In the test generator, we divide operations into three categories according to the number of data sets processed by these operations: element operation, single-set operation, and double-set operation.

**Workload patterns** are designed to combine operations to form complex processing tasks. In the test generator, we abstract three workload patterns: (1) a single-operation pattern contains one single operation; (2) a multi-operation pattern; and, (3) an iterative-operation pattern. The difference between a multi-operation pattern and an iterative-operation pattern is that the former pattern contains finite number of operations, while the latter pattern only provides stopping conditions that the exact number of operations can be known at run time.

**A prescription** includes the information needed to produce a benchmarking test, including data sets, a set of operations and workload patterns, a method to generate workload, and the evaluation metrics.

Figure 4 shows the process of generating a test. At steps 1, 2, and 3, a data set, a set of abstracted operations, and a set of workload patterns are selected, respectively. A prescription is then generated at step 4. Finally, at step 5, a prescribed test for a specific system and software stack is created based on the prescription and system configuration tools. Using the test generator, the workloads in different application domains can be automatically generated.

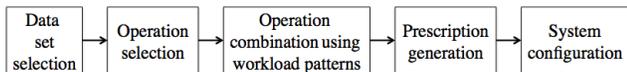

**Figure 4. The benchmark test generation process**

## 4. State-of-the-Art

In this section, we review related work on big data benchmarks from the perspectives of data generation and benchmarking techniques.

### 4.1 Data Generation Techniques

As shown in Table 1, we now review data generation techniques in existing big data benchmarks according to the 4V properties of big data.

**Volume**. To date, most of existing benchmarks generate synthetic data as their workload inputs, where the volume of synthetic data is "scalable". By contrast, some benchmarks such as Hibench and LinkBench also use fixed-size data as inputs. Hence we call these benchmarks "partially scalable" in terms of data volume.

**Velocity**. Some benchmarks such as BigBench, LinkBench, and BigDataBench provide parallel strategies to support the deployment of multiple data generators. In these benchmarks, the data generation rate can be controlled. However, the equally important aspect of data velocity, the data updating frequency, is not considered in these benchmarks. Hence we call these benchmarks "semi-controllable" in terms of data velocity. We also called benchmarks "un-controllable" if both the data generation rate and updating frequency are not considered.

**Variety**. Table 1 lists the data sources of each benchmark's tested data, including tables (structured data); text, graph, and videos (unstructured data); and web logs and resumes (semi-structured data). We can observe that many current benchmarks only consider limited data types (e.g. the text data in Hibench or the table data in YCSB and TPC-DS). Although BigBench and CloudSuite benchmarks support a variety of data sources and types, they are only designed to test applications running in cloud service architecture, and DBMSs and MapReduce Hadoop, respectively.

**Veracity**. In GridMix, PigMix, YCSB, and Micro benchmark, the generation process of synthetic data is independent of the benchmarking applications. For example, in HiBench [12], the synthetic data sets are either randomly generated using the programs in the Hadoop distribution or created using some statistic distributions. Data veracity is "un-considered" in these benchmarks.

In TPC-DS, BigBench, LinkBench, and CloudSuite, the data generation tools partially consider the data veracity. For example, TPC-DS [11] implements a multi-dimensional data generator (MUDD). MUDD generates most of data using traditional synthetic distributions such as a Gaussian distribution. On the other hand, MUDD generates a small portion of crucial data sets using more realistic distributions derived from real data. In BigBench [11], table data are generated using PDGF [16], while web logs and reviews are generated on the basis of the table data. Hence the veracity of web logs and reviews rely on the table data.

By contrast, in BigDataBench [19], different data models are employ to capture and preserve the important characteristics of real data of different types (e.g. table, text, and table). The synthetic data are then generated using the constructed data model, thus avoiding the loss of data veracity.

In conclusion, the issues relating to keeping the 4V properties of big data have not been adequately addressed by current big data benchmarks [1; 3; 4; 6; 7; 9-13; 15; 17].

**Table 1.** Comparison of data generation techniques in existing big data benchmarks.

| Benchmark efforts | Volume | Velocity | Variety (data sources) | Veracity |
|---|---|---|---|---|
| Hibench [12] | Partially scalable | Un-controllable | Texts | Un-considered |
| GridMix [4] | Scalable | Un-controllable | Texts | Un-considered |
| PigMix [6] | Scalable | Un-controllable | Texts | Un-considered |
| YCSB [9] | Scalable | Un-controllable | Tables | Un-considered |
| Performance benchmark [15] | Scalable | Un-controllable | Tables, texts | Un-considered |
| TPC-DS [11] | Scalable | Semi-controllable | Tables | Partially Considered |
| BigBench [11] | Scalable | Semi-controllable | Texts, web logs tables | Partially Considered |
| LinkBench [17] | Partially scalable | Semi-controllable | Graphs | Partially Considered |
| CloudSuite [10] | Partially scalable | Semi-controllable | Texts, graphs, videos, tables | Partially Considered |
| BigDataBench [19] | Scalable | Semi-controllable | Texts, resumes, graphs, tables | Considered |

### 4.2 Benchmarking Techniques

Most of existing big data benchmarks aims to evaluate specific type of systems or architectures. As listed in Table 2, many benchmarks are developed to test the performance of DBMSs and

Hadoop MapReduce, or compare the performance of both types of systems. Specifically, HiBench [12], GridMix [4] and PigMix [6] are designed to test MapReduce Hadoop systems. The performance benchmark in [15] compare two parallel SQL DBMSs (i.e. DBMS-X and Vertica) with MapReduce systems. TPC-DS is TPC's latest decision support benchmark [2] designed to test the performance of DBMSs in decision support systems. Adopting from TPC-DS by adding a web log generator and a review generator, BigBench aims to test the Teradata Aster DBMS and MapReduce systems [11]. In [3], the benchmark is designed to test four SQL driven systems for managing data, including one database (Redshift), one data warehousing systems (Hive), and two engines (Spark and Impala). LinkBench tests MySQL databases that store Facebook's social graph data, and characterizes the real-world database workloads for social applications [17].

Some other benchmarks target at evaluating NoSQL databases or architectures. Yahoo! Cloud Serving Benchmark (YCSB) benchmark compares two non-relational databases (Cassandra and HBase) against one geographically distributed database (PNUTS) and a traditional relational database (MySQL) [9]. The CloudSuite benchmarks in [10; 13] are implemented to test cloud service architectures. Standard Performance Evaluation Corporation (SPEC) [7] has produced several benchmarks for evaluating workstations and has several server and client benchmarks including a Java business benchmark called SPECjBB, but specific big data benchmarks are not available. SPEC had produced web server benchmarks called SPECweb96, SPECweb99, SPECweb2005, and SPECweb2009, but they have been retired. The SPECjEnterprise 2010 and SPEC jBB benchmarks are the closest to big data/cloud benchmarks in SPEC's suites.

At present, BigDataBench is the only big data benchmark that supports the evaluation of a hybrid of different big data systems. The workloads in BigDataBench cover three fundamental and widely usage scenarios (i.e. micro benchmarks, "Cloud OLTP" workloads, and relational queries workloads) and three major application domains in internet services (i.e. Search Engine, Social Network, and E-commerce).

From the perspective of applications users, Table 2 divides workloads in current big data benchmarks into three categories. (1) Online services: these services are sensitive to the response delay, i.e. the time interval between the arrival and departure moments of a service request. Examples of workloads belonging to this category are typical MapReduce operations such as *sort* and *WordCount*. (2) Offline analytics: these services usually perform complex and time-consuming computations on big data. Examples of workloads for testing offline services are machine learning algorithms such as *k-means clustering* and *naive Bayes classification*. (3) Real-time analytics: application users use these services in an interactive manner; that is, a variety of interactions happen between users and application services. Examples of workloads for these services are relational queries such as *selecting*, *joining*, and *aggregation* of database tables.

## 5. Open Challenges

Considering the emergence of new big data applications and the rapid evolution of big data systems, we believe an incremental and iterative approach is necessary to conduct the investigations on big benchmarks. We now propose some challenges to be addressed to develop successful and efficient big data benchmarks.

**Table 2.** Comparison of benchmarking techniques.

| Benchmark efforts | Workloads | | Software stacks |
|---|---|---|---|
| | Type | Examples | |
| Hibench [12] | Offline analytics | Sort, WordCount, TeraSort, PageRank, K-means, Bayes classification | Hadoop and Hive |
| | Real-time analytics | Nutch Indexing | |
| GridMix [4] | Online services | Sort, sampling a large dataset | Hadoop |
| PigMix [6] | Online services | 12 data queries | Hadoop |
| YCSB [9] | Online services | OLTP (read, write, scan, update) | NoSQL systems |
| Performance benchmark [15] | Online services | Data loading, select, aggregate, join, count URL links | DBMS and Hadoop |
| TPC-DS [11] | Online services | Data loading, queries and maintenance | DBMS |
| BigBench [11] | Online services | Database operations (select, create and drop tables) | DBMS and Hadoop |
| | Offline analytics | K-means, classification | |
| LinkBench [17] | Online services | Simple operations such as select, insert, update, and delete; and association range queries and count queries | DBMS |
| CloudSuite [10] | Online services | YCSB's workloads | NoSQL systems, Hadoop, GraphLab |
| | Offline analytics | Text classification, WordCount | |
| BigDataBench [19] | Online services | Database operations (read, write, scan) | NoSQL systems, DBMS, real-time and offline analytics systems |
| | Offline analytics | Micro Benchmarks (sort, grep, WordCount, CFS); search engine (index, PageRank); social network (K-means, connected components (CC)); e-commerce (collaborative filtering (CF), Naive Bayes) | |
| | Real-time analytics | Relational database query (select, aggregate, join) | |

### 5.1 Data-Centric Big Data Benchmarks

The fundamental problem of big data benchmarks is about how to provide better measurement of systems for processing data with 4V properties, which brings the requirement for data-centric benchmarks.

**Fully controllable data velocity**.

The full control of data velocity has two meanings. First, existing big data benchmarks only consider different data generation rates. Hence different data updating frequencies and processing speeds should be reflected in future big data generators.

Secondly, current data velocity is implemented using parallel strategies; that is, data velocity can be controlled by deploying different numbers of parallel data generators. In contrast, we note that data velocity can be controlled in another way: adjusting the efficiency of the data generation algorithms themselves to control data velocity. For example, a graph data generator can be adjusted to consume more memory resources, thus increasing its data generation speed.

**Metrics to evaluate data veracity**.

As discussed in Section 3.2, applying data models to capture and preserve important characteristics of real data is an efficient way to keep data veracity in synthetic data generation. However, how to measure the conformity of the generated synthetic data to the raw data is still an open question; that is, metrics need to be developed to evaluate data veracity. Two types of evaluation metrics can be developed: (1) metrics to compare the raw data and the constructed data models; (2) metrics to compare the raw data and the synthetic data.

This problem is compounded when considering different data sources and data types. For example, to compare a real text data set and a synthetic data, we first need to derive the topic and word distributions from these data sets. Next, statistical metrics such as Kullback–Leibler divergence can be applied to compare the similarity between two distributions. Furthermore, when considering table, graph or even stream data, some other metrics should be developed.

### 5.2 Domain-Centric Big Data Benchmarks

The fast development of big data systems has lead to a number of successful application domains such as scientific analytics, social network, and streaming process. Each of these application domain is the focus of one or multiple big data platform efforts. Domain-centric benchmarks, therefore, are needed to promote the progress of these big data platforms.

**Enriching workloads of big data benchmarks.**

At present, there are three major problems that restrict the wide application of current big data benchmarks. First, there are still many important big data systems such as multimedia systems and applications such as large-scale deep learning algorithms not being considered. Second, in an application domain, a representative workload should reflect both typical data processing operations and the arrival patterns of these operations (i.e. the arriving rate and sequence of operations). We believe profiling history logs of real applications is a good way to obtain the representative arrival patterns. Finally, the truly hybrid workload, i.e. the workload consists of the mix of various data processing operations and their arriving rates and sequences, has not been adequately supported. That is, to the best of our knowledge, none of exiting big data benchmarks is ready to declare itself to be a truly representative and comprehensive big data benchmark until its workloads are significantly enriched to solve the above problems.

**Supporting heterogeneous hardware platforms.**

With the fast development of technology, the emerged hardware platforms and systems significantly change the way about how to process data and show a promising prospect to improve processing efficiency. For example, the heterogeneous platforms of Xeon+General-purpose computing on graphics processing units (GPGPU) and Xeon+Many Integrated Core (MIC) can significantly improve the processing speed of HPC applications. However, to date, both platforms are only limited to the HPC area; that is, the diversity of big data applications are not fully considered in these platforms. For such an issue, big data benchmarks should be developed to evaluate and compare different workloads in state-of-the-practice heterogeneous platforms. The evaluation result is expected to show: (1) whether any platform can consistently win in terms of both performance and energy efficiency for all big data applications, and (2) for each class of big data applications, we hope to find some specific platform that can realize better performance and energy efficiency for them. To support the evaluation of an application, current big data benchmarks should be extended to provide a uniform interface to enable this application running in different platforms. In order to perform apples-to-apples comparisons, this application should be running in the same software stack.

**A reusable environment to automate test generation**.

Section 3.3 proposes a framework to abstract operations and workload patterns from typical data processing behaviors, thus enabling automatic generation of tests with comprehensive workloads for big data systems. We note that these operations and patterns are easy to derive in some application domains. For example, in the application domain of basic database operations, there are some obvious operations such as real, write, select, and delete, and the patterns used to combine these operations are simple. However, in some application domains such as social network, there are a large number of data processing operations such as k-means clustering and collaborative filtering, and the relationships between these operations are complex. All these facts mean abstracting a comprehensive set of operations and behaviors is difficult.

Moreover, in practice, generating benchmarking tests from operations and patterns may be beyond the capabilities of the average system owner. Hence going mainstream with this framework requires the development of an environment that provides abstracted operations and workload patterns for different application domains, as well as offers a repository of reusable prescriptions to simplify the generation of prescribed tests running on state-of-the-art software stacks.

## 6. Conclusion

With the rapid development of information technology, big data systems have emerged to manage and process data with high volume, velocity, and variety. These new systems have given rise to various new requirements about how to develop a new generation of big data benchmarks. In this paper, we summarize the lessons we have learned and propose key challenges in developing big data benchmarks from two aspects: (1) how to develop data generators capable of preserving the 4V properties of big data in data generation; (2) how to automatically generate benchmarking tests to cover a diversity of typical application scenarios while supporting different system implementations and software stacks. We then introduce the methodology on big data benchmark aiming to address the proposed challenges. Next, we discuss existing benchmarking techniques and propose some future research directions. The work presented in this paper represents our effort towards building a truly representative and comprehensive big data benchmark suite and we encourage more investigations and developments in big data benchmarking tools.